\documentclass[12pt]{extarticle}
\pdfoutput=1 
\usepackage{slashed}
\usepackage{latexsym}
\usepackage{amssymb}
\usepackage{amsmath}
\usepackage{autobreak}
\usepackage{graphicx}
\usepackage{arydshln}
\usepackage{adjustbox}
\usepackage{easybmat}
\usepackage{bbm}
\usepackage{cite}
\usepackage{slashed}
\usepackage{bm}
\usepackage{adjustbox}
\usepackage{booktabs}
\usepackage{multirow} 
\usepackage{rotating}
\usepackage{mathtools}
\usepackage{lscape}
\usepackage{hyperref}
\usepackage{bm}
\usepackage{color}
\usepackage{fancybox,framed}
\usepackage{lscape}
\usepackage{multirow}
\usepackage{fancyhdr}
\usepackage{enumitem}

\usepackage[most]{tcolorbox} 

\definecolor{block-gray}{gray}{0.95}

\definecolor{shadecolor}{rgb}{0.01,0.199,0.1}

\newcommand{\highlight}[1]{\colorbox{gray!40}{$\displaystyle #1$}}

\newcommand{\blue}[1]{\textcolor{blue}{#1}}

\newtcolorbox{codeSyntax}{
    enhanced,
    frame hidden,
    colback=block-gray,
    boxrule=0pt,
    borderline west={2pt}{0pt}{gray!80!black}
}

\newcommand{\ii}{\ensuremath{\mathrm{i}}}

\allowdisplaybreaks
\begin{document}
\thispagestyle{empty}
\begin{flushright}
\end{flushright}
\vspace{0.8cm}

\begin{center}
{\Large\sc Renormalization of the SMEFT to dimension\\[0.5cm] eight: Fermionic interactions I}
\vspace{0.8cm}

\textbf{
S. D. Bakshi$^{1,2}$, M. Chala$^2$, \'A. D\'iaz-Carmona$^2$, Z. Ren$^2$ and F. Vilches$^2$
}\\
\vspace{1.cm}
{\em {$^1$ HEP Division, Argonne National Laboratory, Argonne, Illinois 60439, USA}}\\
\vspace{0.3cm}
{\em {$^2$ Departamento de F\'isica Te\'orica y del Cosmos,
Universidad de Granada, Campus de Fuentenueva, E--18071 Granada, Spain}}
\vspace{0.5cm}
\end{center}
\begin{abstract} 
This is the third of a series of works~\cite{Chala:2021pll,DasBakshi:2022mwk} aimed at renormalizing the Standard Model effective field theory at one loop and to order $1/\Lambda^4$, with $\Lambda$ being the new physics cut-off. On this occasion, we concentrate on the running of two-fermion operators induced by pairs of dimension-six interactions. We work mostly off-shell, for which we obtain and provide a new and explicitly hermitian basis of dimension-eight Green's functions.
All our results can be accessed in \href{https://github.com/SMEFT-Dimension8-RGEs}{github.com/SMEFTDimension8-RGEs}. 

\end{abstract}

\newpage


\section{Introduction}
The Standard Model (SM) extended with effective operators, also known as SMEFT~\cite{Buchmuller:1985jz,Grzadkowski:2010es,Brivio:2017vri,Isidori:2023pyp}, is arguably one of the best rivals of the SM itself (if not the only one)  for describing fundamental particles and their interactions at current accessible energies. In order to determine which one accounts best for the experimental data, important progress is needed in both the experimental and the theoretical sides. In this latter respect, SMEFT calculations should be pushed to higher accuracy. This includes computation of observables to $\mathcal{O}(1/\Lambda^2)$ (equivalently dimension-six) at one loop~\cite{Hartmann:2015aia,Hartmann:2016pil,Dawson:2018dxp,Degrande:2020evl,Cullen:2020zof,Corbett:2021iob,Dawson:2021ofa,Alasfar:2022zyr}, as well as to $\mathcal{O}(1/\Lambda^4)$ (equivalently dimension-eight) at tree level~\cite{Corbett:2021eux,Gu:2023emi,Corbett:2021iob,Degrande:2023iob,Ardu:2021koz,Boughezal:2023nhe,Corbett:2023qtg,Martin:2023tvi,Dawson:2024ozw,Li:2024iyj}; where $\Lambda$ is the SMEFT cut-off. Likewise, for consistency as well as for testing the SMEFT against data obtained at very different scales, the dimension-eight SMEFT should be renormalized to the one-loop level.  This paper focuses on this problem. (Ideally, too, the renormalization of the dimension-six sector should be pushed to the two-loop order; see Refs.~\cite{Bern:2020ikv,Jenkins:2023bls,Fuentes-Martin:2023ljp,DiNoi:2024ajj}.)

The first systematic effort towards computing the one-loop SMEFT RGEs to $\mathcal{O}(1/\Lambda^4)$ was initiated in Ref.~\cite{Chala:2021pll}, where quantum corrections to bosonic operators driven by pairs of dimension-six terms were computed. The contributions from dimension-eight terms were later obtained in Ref.~\cite{DasBakshi:2022mwk}. In the meantime, using amplitude methods, the full dimension-eight sector of the SMEFT was renormalized to order $g^2$ in SM couplings, and ignoring flavour, in Ref.~\cite{AccettulliHuber:2021uoa}. Further works, based on the geoSMEFT~\cite{Helset:2022pde,Assi:2023zid}, have also computed a big part of these RGEs (and certain others), with excellent agreement within the different approaches; see also Refs.~\cite{Ardu:2022pzk,Asteriadis:2022ras,Boughezal:2024zqa,Liao:2024xel}. 
These results have been used not only in phenomenological studies~\cite{Durieux:2022hbu,Ardu:2022pzk,Asteriadis:2022ras,Grojean:2024tcw}, but also for understanding positivity constraints better~\cite{Chala:2021wpj,Li:2022aby,Chala:2023jyx,Chala:2023xjy,Ye:2024rzr} as well as a proving ground for high-energy physics tools~\cite{Carmona:2021xtq,Aebischer:2023nnv}. In this work, we continue this endeavour, concentrating on the computation of two-fermion RGEs triggered by pairs of dimension-six terms.

Our approach relies on diagrammatic techniques off-shell. For this matter, we build a new basis of dimension-eight Green's functions with two-fermions and two or more Higgs fields, based on the method for constructing operator bases of Ref.~\cite{Li:2020gnx,Li:2022tec} as well as on the Green's basis of Ref.~\cite{Ren:2022tvi} (which in turn extends the results of Ref.~\cite{Chala:2021cgt}). We further reduce the interactions that are redundant on-shell using a variety of techniques, most importantly diagrammatic on-shell matching relying on the equivalence of the S-matrices computed within the redundant and non-redundant Lagrangians, as described in Refs~\cite{Aebischer:2023nnv,Chala:2024xyz}. To a lesser extent, we also use equations of motion (EoM) both by hand~\cite{Barzinji:2018xvu} and as implemented in the computer tool \texttt{Matchete}~\cite{Fuentes-Martin:2022jrf}, as well as the method briefly introduced in Ref.~\cite{Li:2023cwy} where the off-shell amplitude formalism was proposed, allowing operators to be represented in terms of so-called off-shell amplitudes and to be reduced systematically.

This article is organised as follows. We introduce our conventions in section~\ref{sec:conventions}. In section~\ref{sec:theory} we explain our approach to renormalization, making emphasis on the different classes of operators that run on-shell, as well as on how we build on previous results. We discuss generic features about the structure of mixing under running in section~\ref{sec:rges}, though the full RGEs can be only found in \href{https://github.com/SMEFT-Dimension8-RGEs}{github.com/SMEFTDimension8-RGEs}. We close in section~\ref{sec:conclusions}, where we also reflect on the interplay between our results and positivity constraints, and comment on possible future research directions. In appendix~\ref{app:example}, we provide a detailed example of one of our calculations.

\section{Conventions}
\label{sec:conventions}
Consistently with our previous works~\cite{Chala:2021pll,DasBakshi:2022mwk}, we write the SM Lagrangian as
\begin{align}\nonumber
 \mathcal{L}_\text{SM} = & -\frac{1}{4}G_{\mu\nu}^{A}G^{A\,\mu\nu} -\frac{1}{4}W_{\mu\nu}^{a}W^{a\,\mu\nu} -\frac{1}{4}B_{\mu\nu}B^{\mu\nu}\\\nonumber
 &
+\overline{q_{L}^{\alpha}}\ii\slashed{D}q_{L}^{\alpha}
+\overline{l_{L}^{\alpha}}\ii\slashed{D}l_{L}^{\alpha}
+\overline{u_{R}^{\alpha}}\ii\slashed{D}u_{R}^{\alpha}
+\overline{d_{R}^{\alpha}}\ii\slashed{D}d_{R}^{\alpha}
+\overline{e_{R}^{\alpha}}\ii\slashed{D}e_{R}^{\alpha}
\\
& +\left(D_{\mu}\phi\right)^{\dagger}\left(D^{\mu}\phi\right)
+\mu^{2}|\phi|^{2}-\lambda|\phi|^{4}
-\left(
y_{\alpha\beta}^{u}\overline{q_{L}^{\alpha}}\widetilde{\phi}u_{R}^{\beta}
+y_{\alpha\beta}^{d}\overline{q_{L}^{\alpha}}\phi d_{R}^{\beta}
+y_{\alpha\beta}^{e}\overline{l_{L}^{\alpha}}\phi e_{R}^{\beta}
+\text{h.c.}\right)~,
\end{align}
with minus-sign covariant derivative
$D_\mu = \partial_\mu - \ii g_1 Y B_\mu -\ii g_2\frac{\sigma^I}{2} W_\mu^I -\ii g_s\frac{\lambda^A}{2} G_\mu^A\,$.
%

As usual, $B,W$ and $G$ represent the electroweak gauge bosons and the gluon, respectively, while $g_1,g_2$ and $g_s$ stand for the corresponding gauge couplings. Likewise, $l,q$ and $u,d,e$ are the left-handed leptons and quarks and the right-handed counterparts, respectively. We use $\phi$ for the Higgs doublet; $\sigma^I$ and $\lambda^A$ represent the Pauli and Gell-Mann matrices, respectively.

Hence, ignoring lepton-number violating terms, the SMEFT Lagrangian to order $1/\Lambda^4$ reads:
\begin{align}
 \mathcal{L}_\text{SMEFT} = \mathcal{L}_\text{SM} + \frac{1}{\Lambda^2}\sum_i c_i^{(6)} \mathcal{O}_i^{(6)} + \frac{1}{\Lambda^4}\sum_j c_j^{(8)}\mathcal{O}_j^{(8)}\,.
\end{align}
In our basis, $i$ runs over the dimension-six operators in the Warsaw basis~\cite{Grzadkowski:2010es}, while for $j$ we use that of Ref.~\cite{Murphy:2020rsh}. We follow the notation provided in these references for the corresponding Wilson coefficients (WC).

The dimension-eight WCs vary under changes of the renormalization scale $\tilde{\mu}$ according to:
\begin{equation}\label{eq:gprime}
\dot{c}_i^{(8)} \equiv 16\pi^2\tilde{\mu} \frac{d c_{i}^{(8)}}{d\tilde{\mu}} = \gamma_{ij} c_j^{(8)} + \gamma_{ijk}' c_j^{(6)}c_k^{(6)}\,. 
\end{equation}
The first term in the RHS of this equation amounts to the running induced by dimension-eight terms themselves, while the second term captures the contributions form loops involving pairs of dimension-six terms.

For bosonic operators, both the anomalous dimensions $\gamma$~\cite{DasBakshi:2022mwk} and $\gamma'$~\cite{Chala:2021pll} are known; see also Refs.~\cite{AccettulliHuber:2021uoa,Helset:2022pde,Assi:2023zid}. For two-fermion interactions, $\gamma$ has been computed in Ref.~\cite{AccettulliHuber:2021uoa} to $\mathcal{O}(g^2,\lambda)$ in amplitudes language, while essentially nothing is known about $\gamma'$. The main goal of this paper is computing this quantity. We also compute the $1/\Lambda^4$ corrections to the running of lower-dimensional operators, which scale with powers of the Higgs squared mass parameter $\mu^2$.

We take into account only loops involving dimension-six interactions that can appear at tree level in UV completions of the SMEFT~\cite{Craig:2019wmo,Grojean:2024tcw}. Since, contrary to those in Ref.~\cite{Murphy:2020rsh}, the names of the dimension-six WC  do not shed immediate light on their field content, we summarise them in Tab.~\ref{tab:operators}.

\begin{table}[t!]
 \begin{center}
  \resizebox{0.85\textwidth}{!}{
  \begin{tabular}{cclcl}
   \toprule
   & \textbf{Operator} & \textbf{Notation} & \textbf{Operator} & \textbf{Notation}\\[0.5cm]
   \boldmath{$\phi^6$} & $( \phi^\dagger\phi)^3$ & $\mathcal{O}_{\phi}$ & &\\[0.1cm]
   \boldmath{$\phi^4 D^2$} &  $(\phi^\dagger\phi)\Box(\phi^\dagger\phi)$ & $\mathcal{O}_{\phi\Box}$  &  $(D_\mu\phi^\dagger\phi)(\phi^\dagger D_\mu\phi)$ & $\mathcal{O}_{\phi D}$ \\[0.1cm]
   \hline\\
   \multirow{6}{*}{\boldmath{$\psi^2 \phi^2 D$}} & $\ii(\overline{q}\gamma^\mu q)(\phi^\dagger \overleftrightarrow{D}_\mu\phi)$ & $\mathcal{O}_{\phi q}^{(1)}$ & $\ii(\overline{q}\gamma^\mu q)(\phi^\dagger\sigma^I \overleftrightarrow{D}_\mu^I\phi)$ & $\mathcal{O}_{\phi q}^{(3)}$ \\[0.1cm]
   & $\ii(\overline{u}\gamma^\mu u)(\phi^\dagger \overleftrightarrow{D}_\mu\phi)$ & $\mathcal{O}_{\phi u}$ & $\ii(\overline{d}\gamma^\mu d)(\phi^\dagger \overleftrightarrow{D}_\mu\phi)$ & $\mathcal{O}_{\phi d}$  \\[0.1cm]
   & $\ii(\overline{u}\gamma^\mu d)(\tilde{\phi}^\dagger {D}_\mu\phi)$ & $\mathcal{O}_{\phi ud}$\\[0.1cm]
   & $\ii(\overline{l}\gamma^\mu l)(\phi^\dagger \overleftrightarrow{D}_\mu\phi)$ & $\mathcal{O}_{\phi l}^{(1)}$ & $\ii(\overline{l}\gamma^\mu\sigma^I l)(\phi^\dagger \overleftrightarrow{D}_\mu^I\phi)$ & $\mathcal{O}_{\phi l}^{(3)}$\\[0.1cm]
   & $\ii(\overline{e}\gamma^\mu e)(\phi^\dagger \overleftrightarrow{D}_\mu\phi)$ & $\mathcal{O}_{\phi e}$\\[0.1cm]
   \hline\\
   \multirow{2}{*}{\boldmath{$\psi^2 \phi^3$}} & $(\overline{q}\tilde{\phi}u)\phi^\dagger\phi$ & $\mathcal{O}_{u\phi}$ & $(\overline{q}\phi d)\phi^\dagger\phi$ & $\mathcal{O}_{d\phi}$\\[0.1cm]
   & $(\overline{l}\phi e)\phi^\dagger\phi$ & $\mathcal{O}_{e\phi}$ &  \\[0.1cm]
   \hline\\
   \multirow{14}{*}{\boldmath{$\psi^4$}} & $(\overline{q}\gamma^\mu q)(\overline{q}\gamma_\mu q)$ & $\mathcal{O}_{qq}^{(1)}$ & $(\overline{q}\gamma^\mu\sigma^I q)(\overline{q}\gamma_\mu\sigma^I q)$ & $\mathcal{O}_{qq}^{(3)}$\\[0.1cm]
   & $(\overline{u}\gamma^\mu u)(\overline{u}\gamma_\mu u)$ & $\mathcal{O}_{uu}$ & $(\overline{d}\gamma^\mu d)(\overline{d}\gamma_\mu d)$ & $\mathcal{O}_{dd}$\\[0.1cm]
   & $(\overline{u}\gamma^\mu u)(\overline{d}\gamma_\mu d)$ & $\mathcal{O}_{ud}^{(1)}$ & $(\overline{u}\gamma^\mu T^A u)(\overline{d}\gamma_\mu T^A d)$ & $\mathcal{O}_{ud}^{(8)}$\\[0.1cm]
   & $(\overline{q}\gamma^\mu q)(\overline{u}\gamma_\mu u)$ & $\mathcal{O}_{qu}^{(1)}$ & $(\overline{q}\gamma^\mu T^A q)(\overline{u}\gamma_\mu T^A u)$ & $\mathcal{O}_{qu}^{(3)}$\\[0.1cm]
   & $(\overline{q}\gamma^\mu q)(\overline{d}\gamma_\mu d)$ & $\mathcal{O}_{qd}^{(1)}$ & $(\overline{q}\gamma^\mu T^A q)(\overline{d}\gamma_\mu T^A d)$ & $\mathcal{O}_{qd}^{(8)}$\\[0.1cm]
   & $(\overline{q}u)\epsilon (\overline{q} d)$ & $\mathcal{O}_{quqd}^{(1)}$ & $(\overline{q} T^A u)\epsilon (\overline{q} T^A d)$ & $\mathcal{O}_{quqd}^{(8)}$\\[0.1cm]
   & $(\overline{l}\gamma^\mu l)(\overline{q}\gamma_\mu q)$ & $\mathcal{O}_{lq}^{(1)}$ & $(\overline{l}\gamma^\mu\sigma^I l)(\overline{q}\gamma_\mu \sigma^I q)$ & $\mathcal{O}_{lq}^{(3)}$\\[0.1cm]
   & $(\overline{e}\gamma^\mu e)(\overline{u}\gamma_\mu u)$ & $\mathcal{O}_{eu}$ & $(\overline{e}\gamma^\mu e)(\overline{d}\gamma_\mu d)$ & $\mathcal{O}_{ed}$\\[0.1cm]
   & $(\overline{q}\gamma^\mu q)(\overline{e}\gamma_\mu e)$ & $\mathcal{O}_{qe}$ & $(\overline{l}\gamma^\mu l)(\overline{u}\gamma_\mu u)$ & $\mathcal{O}_{lu}$\\[0.1cm]
   & $(\overline{l}\gamma^\mu l)(\overline{d}\gamma_\mu d)$ & $\mathcal{O}_{ld}$ & $(\overline{l}e)\epsilon (\overline{q}u)$ & $\mathcal{O}_{ledq}$\\[0.1cm]
   & $(\overline{l}e)\epsilon (\overline{q}u)$ & $\mathcal{O}_{lequ}^{(1)}$ & $(\overline{l}\sigma^{\mu\nu}e)\epsilon (\overline{q}\sigma_{\mu\nu}u)$ & $\mathcal{O}_{lequ}^{(3)}$\\[0.1cm]
   & $(\overline{l}\gamma^\mu l)(\overline{l}\gamma_\mu l)$ & $\mathcal{O}_{ll}$ & $(\overline{e}\gamma^\mu e)(\overline{e}\gamma_\mu e)$ & $\mathcal{O}_{ee}$\\[0.1cm]
   & $(\overline{l}\gamma^\mu l)(\overline{e}\gamma_\mu e)$ & $\mathcal{O}_{le}$ &  & \\[0.1cm]
   \bottomrule
  \end{tabular}
  }
 \end{center}
 \caption{\it Dimension-six interactions that can arise at tree level in UV completions of the SMEFT.}\label{tab:operators}
\end{table}

\section{Renormalized Lagrangian}
\label{sec:theory}
\begin{figure}[t]
 \begin{center}
 \includegraphics[width=0.16\columnwidth]{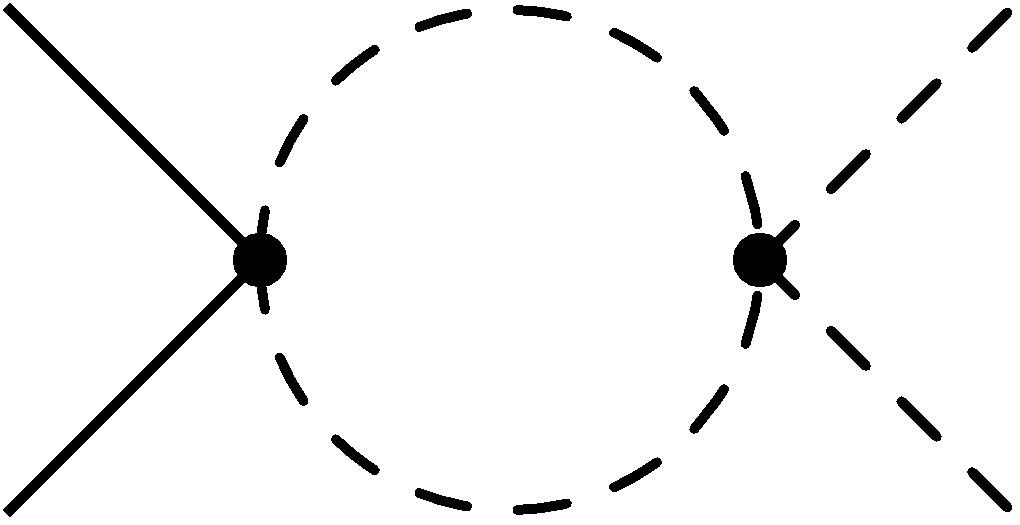} \hfill
 \includegraphics[width=0.16\columnwidth]{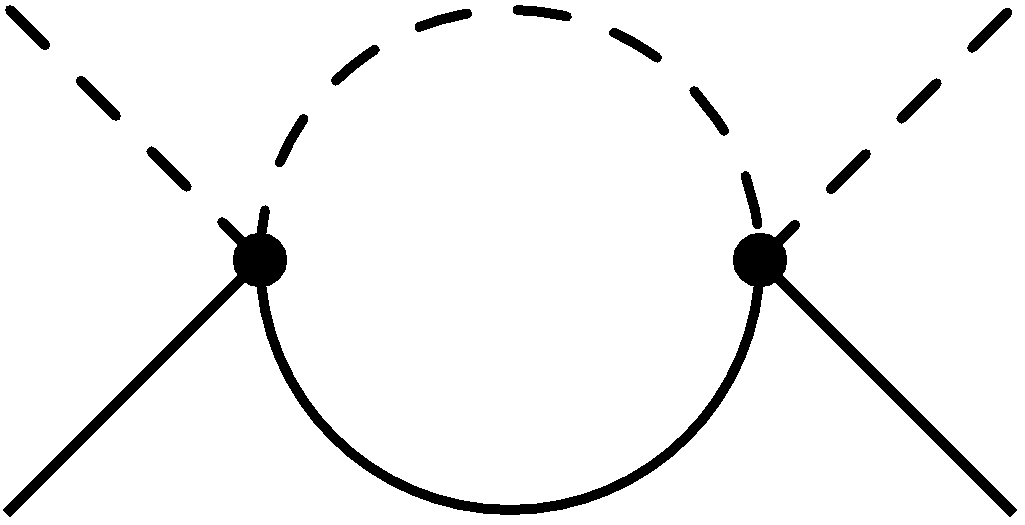} \hfill
 \includegraphics[width=0.16\columnwidth]{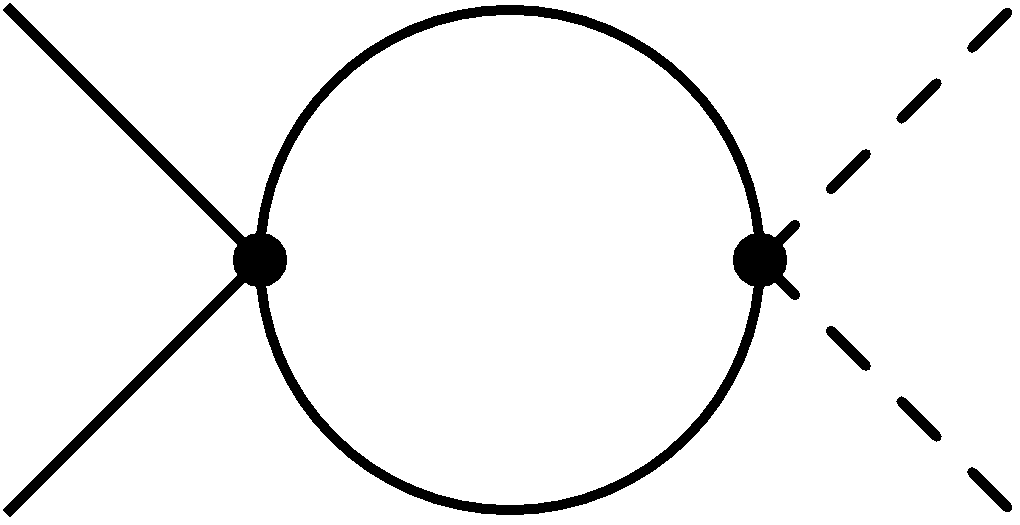} \hfill
 \includegraphics[width=0.16\columnwidth]{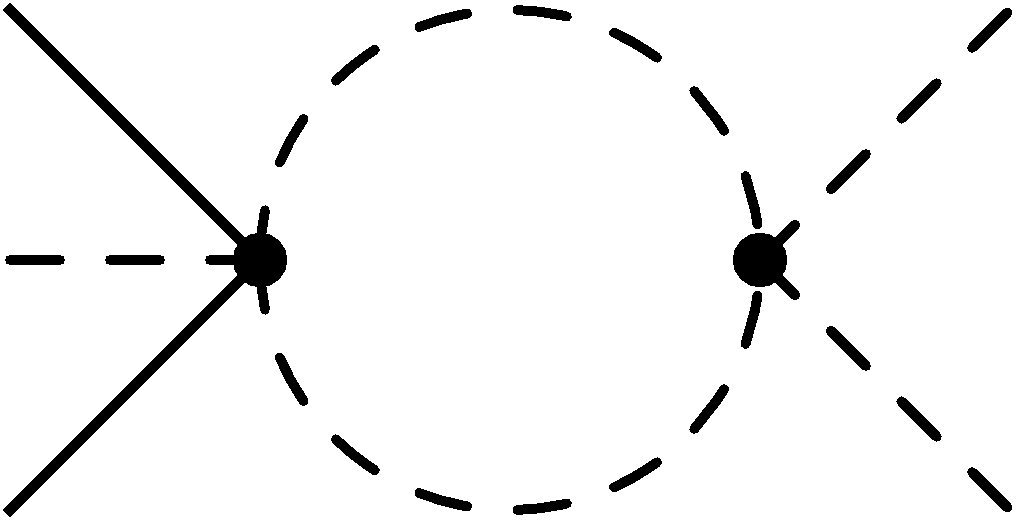} \hfill
 \includegraphics[width=0.16\columnwidth]{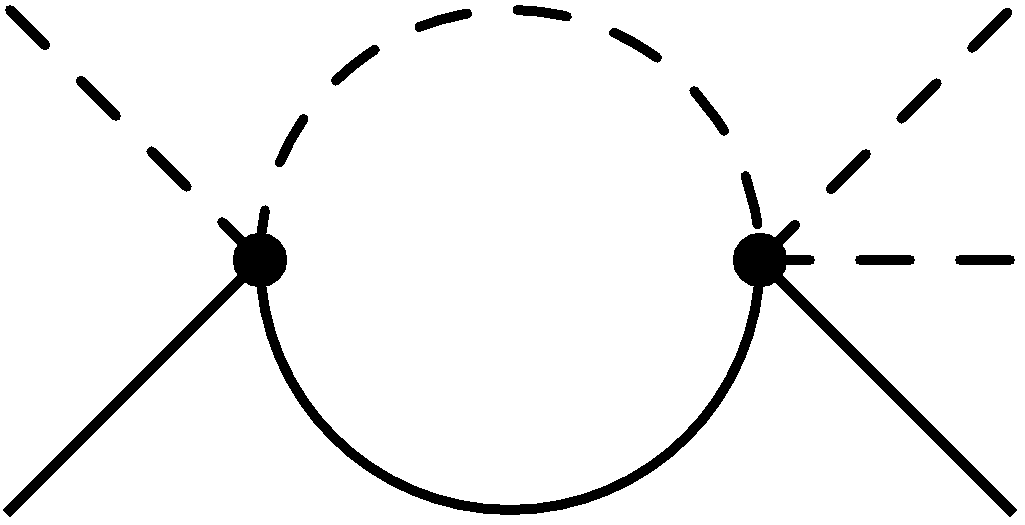}

 \vspace{5mm}

 \hspace{15.8mm}
 \includegraphics[width=0.16\columnwidth]{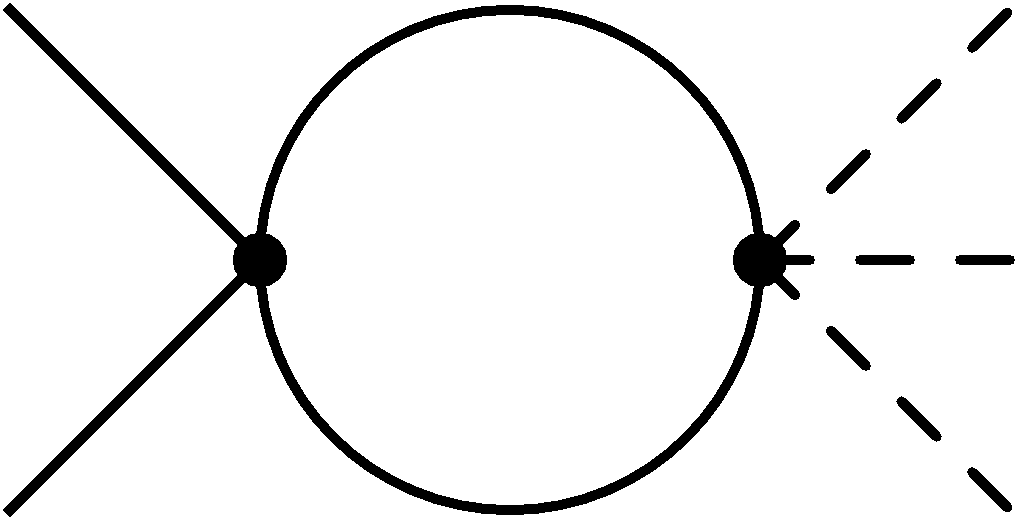}  \hspace{5.4mm}
  \includegraphics[width=0.16\columnwidth]{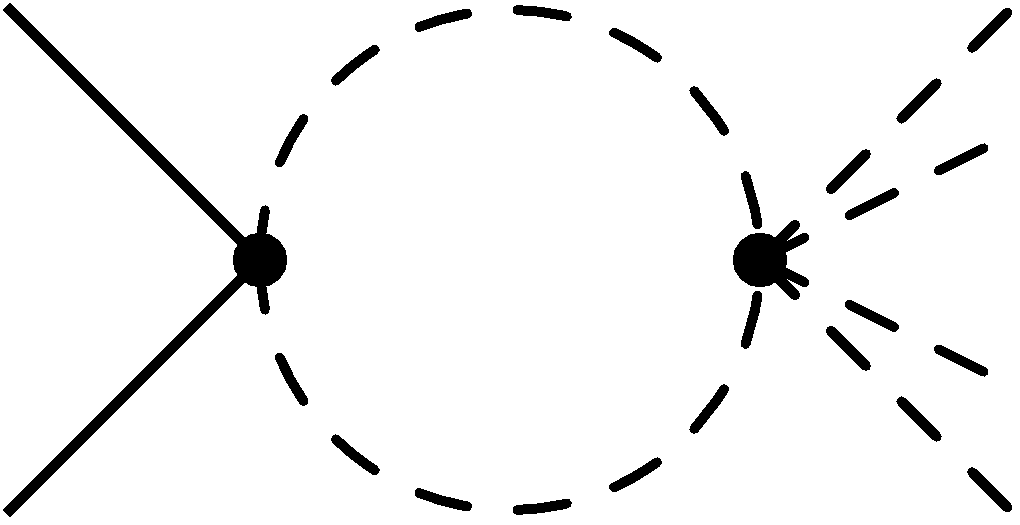} \hspace{5.4mm}
 \includegraphics[width=0.16\columnwidth]{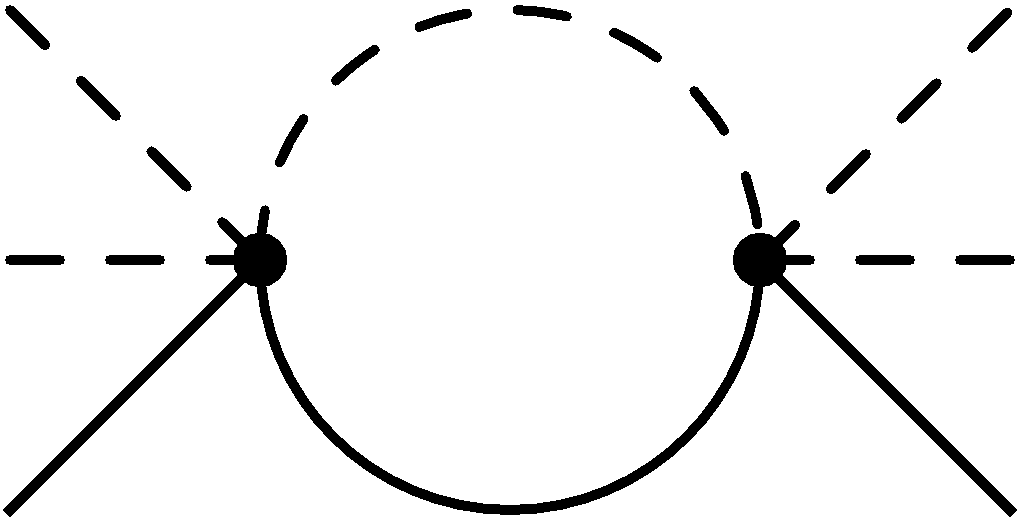} \hspace{5.4mm}
 \includegraphics[width=0.16\columnwidth]{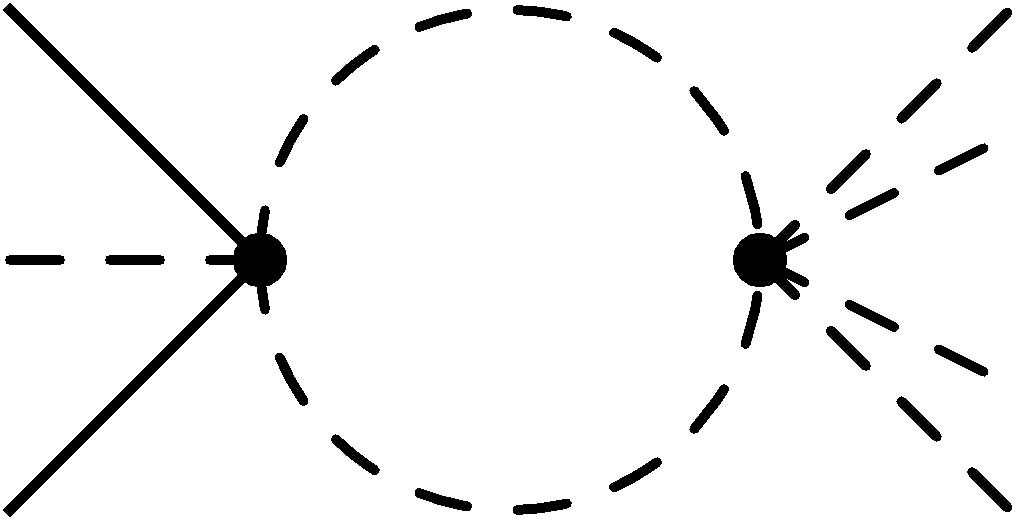} \hspace{15.8mm}

 \end{center}
\caption{\it Two-fermion 1PI topologies (ignoring SM vertices) that arise at one loop involving two dimension-six terms from Tab.~\ref{tab:operators}. \label{fig:topologies}}
\end{figure}
We work in dimensional regularisation with space-time dimension $D=4-2\epsilon$. Our approach to renormalization consists in computing the $1/\epsilon$ poles of all one-loop one-particle-irreducible (1PI) diagrams off-shell, which can be projected onto a Green's basis of effective interactions. For the latter, we use an explicitly-hermitian version of (a subset of) the one presented in Ref.~\cite{Ren:2022tvi} for the dimension-eight sector. We provide this in \texttt{Feynrules}~\cite{Alloul:2013bka} format in \href{https://github.com/SMEFT-Dimension8-RGEs}{github.com/SMEFTDimension8-RGEs}. We have performed our computations using \texttt{FeynArts}~\cite{Hahn:2000kx} and \texttt{FormCalc}~\cite{Hahn:1998yk} as well as within \texttt{matchmakereft}~\cite{Carmona:2021xtq}, with perfect agreement between the two approaches.

The only dimension-eight two-fermion Green's functions that renormalize are those with at least two Higgs fields. This can be drawn from the 1PI topologies that involve two dimension-six terms; see Fig.~\ref{fig:topologies}. However, redundant bosonic dimension-eight Green's functions also contribute to the renormalization of two-fermion dimension-eight operators on-shell. Of those, the only ones that get divergences from loops involving two dimension-six terms are those that contain at least four Higgs fields~\cite{Chala:2021pll}. Since the number of Higgs ($n_\phi$) and of fermion ($n_\psi$) fields in a diagram can only change by $\Delta n_\phi\geq 1,\Delta n_\psi=0$ or by $\Delta n_\phi=-1, \Delta n_\psi=2$ upon attaching a SM vertex on an external leg (equivalently by using EoM on the redundant Green's functions), it is clear that the only two-fermion dimension-eight operators that can renormalize on-shell from dimension-eight bosonic Green's functions contain at least two Higgs fields.
Likewise, redundant dimension-six Green's functions, that we capture using the basis of Ref.~\cite{Gherardi:2020det}, can also contribute to the renormalization of dimension-eight operators on-shell, upon attaching a dimension-six vertex to one external leg.  The only redundant dimension-six Green's functions that renormalize are those in classes~\footnote{Interactions that modify the Higgs or gauge boson propagators are not renormalized off-shell within our setup, because the absence of 3-point vertices within the operators in Tab.~\ref{tab:operators} makes the only possible diagrams being bubbles, which vanish in dimensional regularisation. For the very same reason, Yukawas do not renormalize.} $\phi^4 D^2$, $X\phi^2 D^2$, $\psi^2 \phi D^2$, $\psi^2\phi^2 D$ and $X\psi^2 D$; while the interactions in Tab.~\ref{tab:operators} have $(n_\phi,n_\psi)=(6,0)$, $(4,0)$, $(2,2)$, $(3,2)$ or $(0,4)$.

Accordingly, it is clear that, on-shell, either $\Delta n_\phi\geq 2$, or else $\Delta n_\psi \geq 2$. So once more the only two-fermion dimension-eight operators that can renormalize on-shell from dimension-six Green's functions contain at least two Higgs fields.
Altogether, the only physical two-fermion dimension-eight interactions that can renormalize within our framework are those in classes $\psi^2 \phi^2 D^3$, $X\psi^2\phi^2 D$, $\psi^2\phi^3 D^2$, $X\psi^2\phi^3$, $\psi^2\phi^4 D$ and $\psi^2\phi^5$; for comparison, all two-fermion dimension-six operators in the Warsaw basis do~\cite{Jenkins:2013wua,Jenkins:2013zja,Alonso:2013hga} at order $1/\Lambda^2$ (we find that only some do at order $1/\Lambda^4$). The running of these interactions get contributions from very different Green's functions, at dimensions six and eight, with two or no fermions. The example in Appendix~\ref{app:example} (see in particular Fig.~\ref{fig:example}) should make this transparent.
 
%

\begin{table}[t]
\begin{equation}\nonumber
 %
 %
 %
 \begin{array}{c|c@{\hspace{0.2em}}c@{\hspace{0.2em}}c@{\hspace{0.2em}}c@{\hspace{0.2em}}c@{\hspace{0.2em}}c@{\hspace{0.2em}}c@{\hspace{0.2em}}c}
\highlight{\psi^2\phi^2 D^3} & c_{\phi^4 D^2} & c_{\phi^6} & c_{\psi^2\phi^2 D}  & c_{\psi^2\phi^3} & c_{\psi^4}\\\hline
c_{\phi^4 D^2} & 0 & 0 & \dots & 0 & 0\\
c_{\phi^6} &  & 0 & 0 & 0 & 0\\
c_{\psi^2\phi^2 D} &  &  & \dots & 0 & \blue{\dots}\\
c_{\psi^2\phi^3} &  &  &  & 0 & 0\\
c_{\psi^4} &  &  &  &  & 0\\
\end{array}\\[0.cm]
 \hspace{1cm}
\begin{array}{c|c@{\hspace{0.2em}}c@{\hspace{0.2em}}c@{\hspace{0.2em}}c@{\hspace{0.2em}}c@{\hspace{0.2em}}c@{\hspace{0.2em}}c@{\hspace{0.2em}}c}
\highlight{X \psi^2\phi^2 D} & c_{\phi^4 D^2} & c_{\phi^6} & c_{\psi^2\phi^2 D}  & c_{\psi^2\phi^3} & c_{\psi^4}\\\hline
c_{\phi^4 D^2} & 0 & 0 & g & 0 & 0\\
c_{\phi^6} &  & 0 & 0 & 0 & 0\\
c_{\psi^2\phi^2 D} &  &  & \blue{g} & 0 & \blue{g}\\
c_{\psi^2\phi^3} &  &  &  & 0 & 0\\
c_{\psi^4} &  &  &  &  & 0\\
\end{array}\\[0.cm]
\end{equation}
%
%
\begin{equation}\nonumber
 %
 %
 %
 \begin{array}{c|c@{\hspace{0.2em}}c@{\hspace{0.2em}}c@{\hspace{0.2em}}c@{\hspace{0.2em}}c@{\hspace{0.2em}}c@{\hspace{0.2em}}c@{\hspace{0.2em}}c}
\highlight{\psi^2\phi^3 D^2} & c_{\phi^4 D^2} & c_{\phi^6} & c_{\psi^2\phi^2 D}  & c_{\psi^2\phi^3} & c_{\psi^4}\\\hline
c_{\phi^4 D^2} & \blue{y} & 0 & \blue{y} & \blue{\dots} & \blue{y}\\
c_{\phi^6} &  & 0 & 0 & 0 & 0\\
c_{\psi^2\phi^2 D} &  &  & \blue{y} & \dots & \blue{y}\\
c_{\psi^2\phi^3} &  &  &  & 0 & \blue{\dots}\\
c_{\psi^4} &  &  &  &  & 0\\
\end{array}\\[0.cm]
 \hspace{1cm}
\begin{array}{c|c@{\hspace{0.2em}}c@{\hspace{0.2em}}c@{\hspace{0.2em}}c@{\hspace{0.2em}}c@{\hspace{0.2em}}c@{\hspace{0.2em}}c@{\hspace{0.2em}}c}
\highlight{X \psi^2\phi^3} & c_{\phi^4 D^2} & c_{\phi^6} & c_{\psi^2\phi^2 D}  & c_{\psi^2\phi^3} & c_{\psi^4}\\\hline
c_{\phi^4 D^2} & 0 & 0 & g y & 0 & 0\\
c_{\phi^6} &  & 0 & 0 & 0 & 0\\
c_{\psi^2\phi^2 D} &  &  & g y & g & 0\\
c_{\psi^2\phi^3} &  &  &  & 0 & 0\\
c_{\psi^4} &  &  &  &  & 0\\
\end{array}\\[0.cm]
\end{equation}
%
%
\begin{equation}\nonumber
 %
 %
 %
 \begin{array}{c|c@{\hspace{0.2em}}c@{\hspace{0.2em}}c@{\hspace{0.2em}}c@{\hspace{0.2em}}c@{\hspace{0.2em}}c@{\hspace{0.2em}}c@{\hspace{0.2em}}c}
\highlight{\psi^2\phi^4 D\,\,} & c_{\phi^4 D^2} & c_{\phi^6} & c_{\psi^2\phi^2 D}  & c_{\psi^2\phi^3} & c_{\psi^4}\\\hline
c_{\phi^4 D^2} & y^2 & 0 & \blue{y^2} & y & g^2\\
c_{\phi^6} &  & 0 & 0 & 0 & 0\\
c_{\psi^2\phi^2 D} &  &  & \blue{y^2} & \blue{y} & \blue{y^2}\\
c_{\psi^2\phi^3} &  &  &  & \dots & \blue{y^2}\\
c_{\psi^4} &  &  &  &  & 0\\
\end{array}\\[0.cm]
 \hspace{1cm}
\begin{array}{c|c@{\hspace{0.2em}}c@{\hspace{0.2em}}c@{\hspace{0.2em}}c@{\hspace{0.2em}}c@{\hspace{0.2em}}c@{\hspace{0.2em}}c@{\hspace{0.2em}}c}
\highlight{\,\,\,\psi^2\phi^5\,\,\,\,\,} & c_{\phi^4 D^2} & c_{\phi^6} & c_{\psi^2\phi^2 D}  & c_{\psi^2\phi^3} & c_{\psi^4}\\\hline
c_{\phi^4 D^2} & \blue{y^3} & \blue{y} & \blue{y^3} & \blue{y^2} & \blue{y\lambda}\\
c_{\phi^6} &  & 0 & \blue{y} & \blue{\dots} & \blue{y}\\
c_{\psi^2\phi^2 D} &  &  & y^3 & \blue{y^2} & 0\\
c_{\psi^2\phi^3} &  &  &  & \blue{y} & y^2\\
c_{\psi^4} &  &  &  &  & 0\\
\end{array}\\[0.cm]
\end{equation}
\caption{\it Structure of the mixing of pairs of dimension-six operators into dimension-eight operators. Entries represent the parametric dependence of the leading contributions. Ellipses represent SM-independent contributions. Entries in \blue{blue} are significantly large; see the text for details.}\label{tab:mixing}
\end{table}

\section{Structure of mixing}
\label{sec:rges}
The full set of RGEs for two-fermion dimension-eight operators can be found online at \href{https://github.com/SMEFT-Dimension8-RGEs}{github.com/SMEFTDimension8-RGEs}. For a snapshot, we refer to Tab.~\ref{tab:mixing}, where we show the dominant contributions to the different terms in the anomalous-dimension matrix $\gamma'$, marking in blue those that depart from naive dimensional analysis (namely, $\gamma'\gtrsim 10$ barring SM couplings).

Of the 469 WCs in our Green's basis, 302 renormalize, out of which 184 are redundant. In turn, 114 of the 158 operators in the on-shell basis do~\footnote{The WC $c_{udW\phi^2 D}^{(2)}$ renormalizes only indirectly, through redundant operators; while the WCs $c_{l^2 W\phi^2 D}^{(9)}$, $c_{u^2 G\phi^2 D}^{(4)}$, $c_{d^2 G\phi^2 D}^{(4)}$, $c_{q^2 W\phi^2 D}^{(4)}$ and $c_{q^2 W\phi^2 D}^{(9)}$ renormalize both directly as well as indirectly, but they cancel in the physical basis.}.
Among these, we have the running of dimension-six operators up to $\mathcal{O}(\mu^2/\Lambda^4)$. Given the small number of WCs in here, and for later use, we write explicitly the corresponding RGEs in the lepton sector:
\begin{align}
 \dot{c}_{\phi e,mn} &= 4\mu^2 (c_{\phi\Box} c_{\phi e,mn} + c_{\phi D} c_{\phi e,mn})+\cdots\,,\\
 \dot{c}_{\phi l,mn}^{(1)} &= 4\mu^2(c_{\phi\Box} c_{\phi l,mn}^{(1)} + c_{\phi D} c_{\phi l,mn}^{(1)})+\cdots\,,\\
 \dot{c}_{\phi l,mn}^{(3)} &= 4\mu^2 (c_{\phi l,pn} c_{\phi l,mp}^{(3)} + c_{\phi\Box} c_{\phi l,mn}^{(3)})+\cdots\,,\\
 \dot{c}_{e\phi ,mn} &= -\mu^2\bigg[48 c_{\phi\Box} c_{e\phi,mn} - 12 c_{\phi D} c_{e\phi,mn} +  2 c_{e\phi,m p} c_{\phi e,pn} - 2 c_{e\phi,pn} c_{\phi l,mp}^{(1)}  -6 c_{e\phi,pn} c_{\phi l,mp}^{(3)} \nonumber\\
 & - 8 c_{\phi\Box} c_{\phi e,pn} y_{mp}^e + 2 c_{\phi D} c_{\phi e,pn} y_{mp}^e + 8 c_{\phi\Box} c_{\phi l,mp}^{(1)} y_{pn}^e - 
 2 c_{\phi D} c_{\phi l,mp}^{(1)} y_{p n}^e + 
 24 c_{\phi\Box} c_{\phi l,mp}^{(3)} y_{pn}^e \nonumber\\
 &- 
 6 c_{\phi D} c_{\phi l,mp}^{(3)} y_{pn}^e  +2 c_{\phi e,pn} c_{\phi e,qp} y_{mq}^e - 
 4 c_{\phi e,pn} c_{\phi l,mq}^{(1)} y_{qp}^e  - 4 c_{\phi e,pn} c_{\phi l,mq}^{(3)} y_{qp}^e + 2 c_{\phi l,mp}^{(1)} c_{\phi l,pq}^{(1)} y_{q,n}^e \nonumber\\
 &+ 2 c_{\phi l,mp}^{(1)} c_{\phi l,pq}^{(3)} y_{qn}^e  
 +2 c_{\phi l,pq}^{(1)} c_{\phi l,mp}^{(3)} y_{qn}^e + 
 6 c_{\phi l,mp}^{(3)} c_{\phi l,pq}^{(3)} y_{qn}^e+  12 c_{e\phi,pq} c_{le,mpqn} - 16 c_{\phi\Box} c_{le,mpqn} y_{pq}^e \nonumber\\
 &  + 
 4 c_{\phi D} c_{le,mpqn} y_{pq}^e  + 24 c_{\phi\Box} c_{ledq,mnpq} y_{qp}^d  -  6 c_{\phi D} c_{ledq,mnpq} y_{qp}^d - 24 c_{\phi\Box} c_{lequ,mnpq}^{(1)} y_{pq}^{u*}   \nonumber\\
 &  + 
 6 c_{\phi D} c_{lequ,mnpq}^{(1)} y_{pq}^{u*} - 
 18 c_{d\phi,pq} c_{ledq,mnqp} +  18 c_{lequ,mnpq}^{(1)} c_{u\phi,pq}^* - 
 16 c_{\phi\Box}^2 y_{mn}^e + 10 c_{\phi\Box} c_{\phi D} y_{mn}^e \nonumber\\
 &- 2 c_{\phi D}^2 y_{mn}^e \bigg]+\cdots\,;\label{eq:cephi}
 \end{align}
the ellipses represent terms independent of $\mu^2$, which were first computed in Refs.~\cite{Jenkins:2013wua,Jenkins:2013zja}, and we have omitted the cut-off $\Lambda$.

In appendix~\ref{app:example} we provide a thorough explanation of how we derive Eq.~\eqref{eq:cephi}. Finally, SM dimension-four terms with two fermions, that is Yukawa couplings, do not renormalize to $\mathcal{O}(\mu^4/\Lambda^4)$.

\section{Discussion and outlook}
\label{sec:conclusions}
We have computed the one-loop RGEs of the two-fermion operators of the SMEFT to $\mathcal{O}(1/\Lambda^4)$ triggered by pairs of dimension-six terms. To this aim, we have built on previous results on the bosonic sector~\cite{Chala:2021pll}. The full result can be found in a \texttt{Mathematica} notebook at \href{https://github.com/SMEFT-Dimension8-RGEs}{github.com/SMEFTDimension8-RGEs}. 
Several comments are in order. \\

\textbf{1}. To the best of our knowledge, close to none of these RGEs have been computed anywhere else in the literature. We have cross-checked some minor terms in the RGEs of $c_{e^2\phi^4D}$, $c_{l^2\phi^4 D}^{(1)}$ and $c_{l^2\phi^4 D^2}^{(2)}$ with Ref.~\cite{Ardu:2022pzk}. (Most of the terms entering these RGEs are neglected in this reference because they are not relevant for the flavour-violating observables studied in there.)\\

\textbf{2}. Some further cross-checks ensue from positivity bounds~\cite{Adams:2006sv}. Indeed, as explained in Refs.~\cite{Chala:2021wpj,Chala:2023jyx}, one-loop quantum corrections triggered by pairs of dimension six terms do not spoil the validity of tree-level positivity bounds. In particular, it must be the case that~\cite{Chala:2023xjy} $\dot{c}_{e^2\phi^2 D^3}^{(1)}+\dot{c}_{e^2\phi^2 D^3}^{(2)}\geq0$ as well as $\dot{c}_{ l^2\phi^2 D^3}^{(1)}+\dot{c}_{ l^2\phi^2 D^3}^{(2)}+\dot{c}_{ l^2\phi^2 D^3}^{(3)}+\dot{c}_{ l^2\phi^2 D^3}^{(4)}\geq0$ and $\dot{c}_{ l^2\phi^2 D^3}^{(1)}+\dot{c}_{ l^2\phi^2 D^3}^{(2)}-\dot{c}_{ l^2\phi^2 D^3}^{(3)}-\dot{c}_{ l^2\phi^2 D^3}^{(4)}\geq0$ in the one-flavour limit for arbitrary values of the dimension-six WCs; likewise in the quark sector. In fact, we get:
\begin{align}
 \dot{c}_{e^2\phi^2 D^3}^{(1)}+\dot{c}_{e^2\phi^2 D^3}^{(2)} &= 4 c_{\phi e}^2\,,\\
 \dot{c}_{ l^2\phi^2 D^3}^{(1)}+\dot{c}_{ l^2\phi^2 D^3}^{(2)}+\dot{c}_{ l^2\phi^2 D^3}^{(3)}+\dot{c}_{ l^2\phi^2 D^3}^{(4)} &= 4 (c_{\phi l}^{(1)}+c_{\phi l}^{(3)})^2+8 (c_{\phi l}^{(3)})^2\,,\\
 \dot{c}_{ l^2\phi^2 D^3}^{(1)}+\dot{c}_{ l^2\phi^2 D^3}^{(2)}-\dot{c}_{ l^2\phi^2 D^3}^{(3)}-\dot{c}_{ l^2\phi^2 D^3}^{(4)} &= 4 (c_{\phi l}^{(1)}-c_{\phi l}^{(3)})^2+8 (c_{\phi l}^{(3)})^2\,.
\end{align}

\textbf{3}. It is self-evident that none of the two-fermion dimension-eight operators that can only arise at loop level in UV completions of the SMEFT renormalizes, simply because they all contain less than one Higgs field. This extends previous findings in the bosonic sector~\cite{Chala:2021pll}. \\

We can mention several obvious future directions of work. First, it remains to compute the renormalization of two-fermion dimension-eight operators by loops involving single dimension-eight terms, as well as completing the running of four-fermion interactions~\cite{Boughezal:2024zqa,Liao:2024xel}. Moreover, our results could be applied to describe, to a better accuracy, the breaking of universality with the SMEFT~\cite{Wells:2015cre} as well as departures from different flavour assumptions~\cite{Greljo:2023bdy} due to quantum corrections. Likewise, it would be interesting to study the impact on electroweak precision parameters, though this requires first knowledge on how the latter are written within our adopted SMEFT basis; see Refs.~\cite{Wells:2015uba,Wells:2015cre,Corbett:2023qtg,Corbett:2024yoy} for some results. Finally, and most importantly, our results pave the way to confronting the SMEFT with experimental data with higher precision. 

Still, given the huge amount of computations at hand (for an example, the full list of redundancies, like the one in Eq.~\eqref{eq:ind}, involves more than 2000 monomials; its computation surpassing in complexity the work needed for obtaining the counter-terms of all relevant WCs), it would be very desirable to cross-check our findings (using same or alternative methods). To facilitate this process, we also provide separate notebooks for the redundancies and for the divergences in the \texttt{GitHub} repository. We will appreciate being informed about any disagreement.

\section*{Acknowledgments}
We would like to thank J.L. Miras for sharing his modified version of \texttt{Feynrules} free of bugs related to effective interactions.  We also thank M. Ramos for comments on the manuscript. We are especially grateful to Jos\'e Santiago for helping us with \texttt{matchmakereft}. This work has been partially funded by MICIU/AEI/10.13039/501100011033 and by the European Union NextGenerationEU/PRTR under grant CNS2022-136024, by ERDF/EU (grants PID2021-128396NB-I00 and PID2022-139466NB-C22), by the Junta de Andaluc\'ia grants FQM 101, P21-00199 as well as by Consejer\'ia de Universidad, Investigaci\'on e Innovaci\'on, Gobierno de España and Uni\'on
Europea -- NextGenerationEU under grant AST22 6.5. SDB, MC and ADC are further supported by DOE contract DE-AC02-06CH11357, by the Ram\'on y Cajal program under grant RYC2019-027155-I, and by the FPI program, respectively.

\appendix

\section{Detailed example: renormalization of $\mathcal{O}_{e\phi}$}
\label{app:example}
The operator $\mathcal{O}_{e\phi}=(\overline{l}\phi e)\phi^\dagger\phi$ renormalizes \textit{directly} through loops like the upper fourth and upper fifth ones in Fig.~\ref{fig:topologies}. The bottom first does not contribute to the running of dimension-six Green's functions because there is no light mass on the loop (fermions are massless in the Higgs unbroken phase). The corresponding contribution reads:
\begin{align}\label{eq:dir}
 (\dot{c}_{e\phi,mn})^\text{dir} &= -\mu^2\bigg[8 (3c_{\phi\Box}-c_{\phi D}) c_{e\phi,mn} + 2 c_{e\phi,mp} c_{\phi e,pn} -2 (c_{\phi l,mp}^{(1)}+3 c_{\phi l,mp}^{(3)}) c_{e\phi,pn} \nonumber\\
 & - 4 c_{\phi D} (c_{\phi l,mp}^{(1)}+c_{\phi l,mp}^{(3)}) y^e_{pn} + 4 c_{\phi D} y^e_{mp} c_{\phi e,pn} - 4 (c_{\phi l,mr}^{(1)}+c_{\phi l,mr}^{(3)}) y^e_{rp} c_{\phi e,pn}\bigg]\,.
\end{align}

\begin{figure}[t]
 \includegraphics[width=\columnwidth]{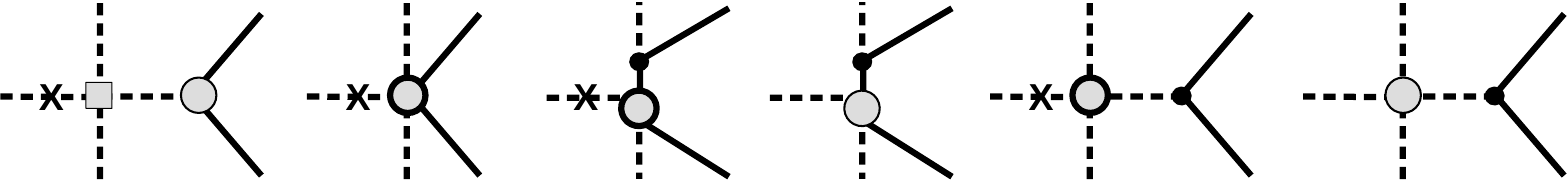}
 \caption{\it Pictorial representation of the different contributions entering the renormalization of $\psi^2\phi^3$. Thin- and thick-line circles $\bigcirc$ represent redundant operators of dimension six and eight, respectively; a square $\square$ stands for physical dimension-six interactions, while the small dots are renormalizable ones. The cross $\times$ represents a Higgs mass insertion.
 }\label{fig:example}
\end{figure}

On top of this, $\mathcal{O}_{e\phi}$ renormalizes also \textit{indirectly} upon using the EoM on redundant Green's functions (equivalently, upon attaching a SMEFT vertex on an external line on a 1PI diagram).  We represent the different possibilities in Fig.~\ref{fig:example}.
The first one describes the contribution from a redundant dimension-six Green's function in the class $\psi^2\phi D^2$ with an EoM involving $\phi^4 D^2$.
The cross represents a mass insertion (equivalently the mass-dependent part of the Higgs EoM $D^2\phi \sim \mu^2\phi$). Note that something interesting occurs here: While, normally, the Higgs EoM  removes two derivatives from the Green's function where it is applied, the Higgs EoM of the SMEFT~\footnote{We use the results in Ref.~\cite{Barzinji:2018xvu}, with typos corrected. In particular, there is a factor of $2$ off in the contribution of $c_{\phi D}$ to the EoM of the $W$ as well as a wrong imaginary unit in the $\epsilon^{IJK}$ term. Note also that the Yukawa and covariant-derivative-sign conventions in that reference differ from ours.} to order $1/\Lambda^2$ includes $D^2$ terms that bring these derivatives back. That is why two EoM (two vertex insertions in the 1PI diagram) are needed to move $\psi^2\phi D^2$ to $\psi^2\phi^3$.

The second stands for the effect of a redundant dimension-eight $\psi^2 \phi^3 D^2$ with a Higgs mass insertion. 
The third describes the contribution from a redundant dimension-eight $\psi^2\phi^2 D^3$ with the Yukawa part of a fermion EoM and a Higgs mass insertion. The fourth represents the direct renormalization of the dimension-six $\psi^2\phi^2 D$ (necessarily ensuing from diagrams with Higgs loops providing a $\mu^2$ term), which turns into $\psi^2\phi^3$ via a Yukawa-like EoM. (No Higgs mass insertion in external legs is needed in this case.)
The last-to-last one stands for the effect of a purely bosonic redundant dimension-eight $\phi^4 D^4$ with the Yukawa part of the Higgs EoM and a Higgs mass insertion. Finally, there are also effects from redundant dimension-six $\phi^4 D^2$ that receive $\mu^2/\Lambda^4$ divergences, shown in the final diagram.

In practice, we compute most of the effects of all aforementioned EoM using tree-level on-shell matching~\cite{Aebischer:2023nnv,Chala:2024xyz}. For dimension-six redundant terms, we also use the dimension-six EoM, which generate a set of redundant dimension-eight interactions, followed again by tree-level on-shell matching.

Altogether, in our Green's basis, the indirect renormalization of $\mathcal{O}_{e\phi}$ reads:
%
\begin{align}\label{eq:ind}
 (\dot{c}_{e\phi,mn})^\text{ind} &= \frac{1}{2} r_{\phi D}' y_{mn}^e 
 +r_{\phi e,pn}' y_{mp}^e 
+ 
 r_{\phi l,mp}^{\prime(1)} y_{pn}^e 
 %
 +r_{\phi l,mp}^{\prime (3)} y_{pn}^e 
-\mu^2 \bigg[-4 c_{\phi\Box} r_{e\phi D,mn}^{(1)}  + c_{\phi D} r_{e\phi D,mn}^{(1)} \nonumber\\
 &- 
 2 c_{\phi\Box} r_{e\phi D,mn}^{(2)}+ \frac{1}{2} c_{\phi D} r_{e\phi D,mn}^{(2)}+ 
 2 c_{\phi\Box} r_{e\phi D,mn}^{(4)}- \frac{1}{2} c_{\phi D} r_{e\phi D,mn}^{(4)}+  r_{le\phi^3 D^2,mn}^{(7)}\nonumber\\
 & + \ii r_{le\phi^3 D^2,mn}^{(9)}+ 
 r_{le\phi^3 D^2,mn}^{(11)} - r_{le\phi^3 D^2,mn}^{(14)}+ \frac{1}{2} r_{le\phi^3 D^2,mn}^{(15)} + \frac{3}{2} \ii r_{le\phi^3 D^2,mn}^{(16)}- 
 r_{\phi^4 D^4}^{(4)} y_{mn}^e \nonumber\\
 & + 2 r_{\phi^4 D^4}^{(8)} y_{mn}^e + 
 r_{\phi^4 D^4}^{(10)} y_{mn}^e  + r_{\phi^4 D^4}^{(11)} y_{mn}^e + \frac{1}{2} r_{l^2\phi^2D^3,mp}^{(33)} y_{pn}^e+ 
 \frac{1}{2} r_{l^2\phi^2D^3,mp}^{(35)} y_{pn}^e\bigg]\,.
 %
 %
\end{align}
Note that we name the redundant WCs with $r$. Let us also stress that, in full generality, there are more redundant WCs that enter this equation, however they are not renormalized within our framework.

We have:
\begin{align}
 r'_{\phi e,mn} &= -2\mu^2 c_{\phi e,mp} c_{\phi e,pn}\,,\\
 %
 %
 r'^{(1)}_{\phi l,mn} &= -2\mu^2 c_{\phi l,mp}^{(1)} c_{\phi l,pn}^{(1)} - 6\mu^2 c_{\phi l,mp}^{(3)} c_{\phi l,pn}^{(3)}\,,\\
 r'^{(3)}_{\phi l,mn} &= -2\mu^2 c_{\phi l,mp}^{(1)} c_{\phi l,pn}^{(3)} - 2\mu^2 c_{\phi l,mp}^{(3)} c_{\phi l,pn}^{(1)}\,,\\
 r_{e\phi D,mn}^{(1)} &= -2 c_{\phi e,pn} y^e_{mp} +4 c_{le,mprn} y^e_{pr} -6 c_{ledq,mnpr} y^d_{rp} +6 c_{lequ,mnpr}^{(1)} y^{u*}_{pr}\,,\\
 r_{e\phi D,mn}^{(2)} &=  c_{\phi e,pn} y^e_{mp}-c^{(1)}_{\phi l,mp} y^e_{pn}-3c^{(3)}_{\phi l,mp} y^e_{pn}  \,,\\
 %
 %
 r_{e\phi D,mn}^{(4)} &= - 3c_{\phi e,pn} y^e_{mp} -c^{(1)}_{\phi l,mp} y^e_{pn}-3c^{(3)}_{\phi l,mp} y^e_{pn} \,,\\
 r_{le\phi^3 D^2,mn}^{(7)} &= 12 c_{\phi\square} c_{e\phi,mn}- 2c_{\phi D} c_{e\phi,mn}-c_{e\phi,pn}c^{(1)}_{\phi l,mp}-5c_{e\phi,pn}c^{(3)}_{\phi l,mp} -4c_{\phi\square} c_{\phi e,pn} y^e_{mp} \nonumber \\ & \quad \,+c_{\phi D} c_{\phi e,pn} y^e_{mp}+4 c_{\phi\square} c^{(1)}_{\phi l,mp} y^e_{pn}+ 16 c_{\phi\square} c^{(3)}_{\phi l,mp} y^e_{pn} - 3 c_{\phi D} c^{(3)}_{\phi l,mp} y^e_{pn}\nonumber \\ & \quad \,- c_{\phi e,pn} c^{(1)}_{\phi l,mr} y^e_{rp}+ 2c_{\phi e,pn} c^{(3)}_{\phi l,mr} y^e_{rp} - c^{(1)}_{\phi l,mp} c^{(3)}_{\phi l,pr} y^e_{rn}+c^{(3)}_{\phi l,pr} c^{(3)}_{\phi l,mp} y^e_{rn} \nonumber \\ & \quad \, +4c_{e\phi,pr} c_{le,mprn}  +4 c^{(3)}_{\phi l,pr} c_{le,mpsn} y^e_{rs} -6c_{d\phi,pr} c_{ledq,mnrp} + 6 c^{(1)}_{lequ,mnpr}c^*_{u\phi,pr}\nonumber \\ & \quad \, -6 c^{(3)}_{\phi q,pr}c_{ledq,mnsp}y^d_{rs} +6 c^{(3)}_{\phi q,pr} c_{lequ,mnrs}^{(1)} y^{u*}_{ps}-3 c_{\phi ud,pr}c^{(1)}_{lequ,mnsp}y^{d*}_{sr} \nonumber \\ & \quad \,  +3 c_{\phi ud,pr} c_{ledq,mnrs} y^u_{sp}  \,,\\
 r_{le\phi^3 D^2,mn}^{(9)} &= i c_{e\phi,mp} c_{\phi e,pn}-ic_{e\phi,pn} c^{(1)}_{\phi l,mp}-6ic_{e\phi,pn} c^{(3)}_{\phi l,mp}+6ic_{\phi\square} c^{(3)}_{\phi l,mp} y^e_{pn} \nonumber \\ & \quad \, -\frac{3}{2}ic_{\phi D} c^{(3)}_{\phi l,mp} y^e_{pn} -6ic_{\phi e,pn} c^{(1)}_{\phi l,mr} y^e_{rp} + 12 ic_{\phi e,pn} c^{(3)}_{\phi l,mr} y^e_{rp}\,,\\
 r_{le\phi^3 D^2,mn}^{(11)} &= 12c_{\phi\square}c_{e\phi,mn} -2 c_{\phi D}c_{e\phi,mn}+c_{e\phi,pn} c^{(1)}_{\phi l,mp}+4c_{e\phi,pn} c^{(3)}_{\phi l,mp}-4c_{\phi\square} c_{\phi e,pn} y^e_{mp}\nonumber \\ & \quad \,-c_{\phi D} c_{\phi e,pn} y^e_{mp} +4c_{\phi\square} c^{(1)}_{\phi l,mp} y^e_{pn}  + 10c_{\phi\square} c^{(3)}_{\phi l,mp} y^e_{pn} +\frac{1}{2} c_{\phi D} c^{(3)}_{\phi l,mp} y^e_{pn} \nonumber \\ & \quad \,  - c_{\phi e,pn} c^{(1)}_{\phi l,mr} y^e_{rp} +2c_{\phi e,pn} c^{(3)}_{\phi l,mr} y^e_{rp}+ c^{(1)}_{\phi l,mp} c^{(3)}_{\phi l,pr} y^e_{rn} - c^{(3)}_{\phi l,mp} c^{(3)}_{\phi l,pr} y^e_{rn} \nonumber \\ & \quad \,+4c_{e\phi,pr} c_{le,mprn} + 4 c^{(3)}_{\phi l,pr} c_{le,mpsn} y^e_{rs}-6c_{d\phi,pr} c_{ledq,mnrp}+6 c^{(1)}_{lequ,mnpr} c^*_{u\phi,pr}\nonumber \\ & \quad \,-6 c^{(3)}_{\phi q,pr} c_{ledq,mnsp} y^d_{rs}+ 6c^{(3)}_{\phi q,pr} c^{(1)}_{lequ,mnrs} y^{u*}_{ps}  -3 c_{\phi ud,pr} c^{(1)}_{lequ,mnsp} y^{d*}_{sr}\nonumber \\ & \quad \, +3c_{\phi ud,pr} c_{ledq,mnrs} y^u_{sp} \,,\\
 r_{le\phi^3 D^2,mn}^{(14)} &= -c_{e\phi,mp} c_{\phi e,pn}- c_{e\phi,pn} c^{(1)}_{\phi l,mp}-7c_{e\phi,pn} c^{(3)}_{\phi l,mp}+2c_{\phi D} c_{\phi e,pn}y^e_{mp}-c_{\phi D}c^{(1)}_{\phi l,mp} y^e_{pn}\nonumber \\ & \quad \, +8c_{\phi\square}c^{(3)}_{\phi l,mp} y^e_{pn}-3c_{\phi D}c^{(3)}_{\phi l,mp} y^e_{pn} -2 c_{\phi e,pn}c^{(1)}_{\phi l,mr} y^e_{rp}+4 c_{\phi e,pn}c^{(3)}_{\phi l,mr} y^e_{rp}\nonumber \\ & \quad \, -2c^{(1)}_{\phi l,mp}c^{(3)}_{\phi l,pr}y^e_{rn} +2c^{(3)}_{\phi l,pr}c^{(3)}_{\phi l,mp}y^e_{rn}-4c_{e\phi,pr} c_{le,mprn} +8c^{(3)}_{\phi l,pr}c_{le,mpsn}y^e_{rs}\nonumber \\ & \quad \,+6c_{d\phi,pr} c_{ledq,mnrp}  -6c^{(1)}_{lequ,mnpr}c^*_{u\phi,pr}  -12c^{(3)}_{\phi q,pr} c_{ledq,mnsp} y^d_{rs}   \nonumber \\ & \quad \,+12c^{(3)}_{\phi q,pr} c^{(1)}_{lequ,mnrs} y^{u*}_{ps}-6 c_{\phi ud,pr} c^{(1)}_{lequ,mnsp} y^{d*}_{sr}  +6c_{\phi ud,pr} c_{ledq,mnrs}y^u_{sp}  \,,\\
 r_{le\phi^3 D^2,mn}^{(15)} &=  -3 c_{e\phi,mp} c_{\phi e,pn} - c_{e\phi,pn} c^{(1)}_{\phi l,mp} - 9c_{e\phi,pn} c^{(3)}_{\phi l,mp} + 2c_{\phi D} c^{(1)}_{\phi l,mp} y^e_{pn} \nonumber \\ & \quad \, +12 c_{\phi \square} c^{(3)}_{\phi l,mp} y^e_{pn} -5 c_{\phi D} c^{(3)}_{\phi l,mp} y^e_{pn} -4 c^{(1)}_{\phi l,mp} c^{(3)}_{\phi l,pr} y^e_{rn}+4 c^{(3)}_{\phi l,mp} c^{(3)}_{\phi l,pr} y^e_{rn}\,,\\
 r_{le\phi^3 D^2,mn}^{(16)} &= -i c_{e\phi,mp} c_{\phi e,pn} + ic_{e\phi,pn} c^{(1)}_{\phi l,mp} + 5ic_{e\phi,pn} c^{(3)}_{\phi l,mp} -4i c_{\phi \square} c^{(3)}_{\phi l,mp} y^e_{pn} \nonumber \\ & \quad \, +i c_{\phi D} c^{(3)}_{\phi l,mp} y^e_{pn} +4i c_{\phi e,pn} c^{(1)}_{\phi l,mr} y^e_{rp}-8i c_{\phi e,pn} c^{(3)}_{\phi l,mr} y^e_{rp}\,,\\
 r_{l^2\phi^2 D^3,mn}^{(33)} &= -2c^{(3)}_{\phi l,pn} c^{(3)}_{\phi l,mp}\,,\\
 r_{l^2\phi^2 D^3,mn}^{(35)} &= 2c^{(3)}_{\phi l,pn} c^{(3)}_{\phi l,mp}\,.
\end{align}

The contributions from bosonic Green's functions were computed previously in Ref.~\cite{Chala:2021pll}, obtaining:
\begin{align}
 r_{\phi D}^\prime &= - \mu^2 (-16 c_{\phi\Box}^2 + 8 c_{\phi\Box} c_{\phi D} + 2 c_{\phi D}^2)\,,\\
 %
 %
 r_{\phi^4 D^4}^{(4)} &= -24 c_{\phi\Box}^2 + 2 c_{\phi\Box} c_{\phi D} + c_{\phi D}^2\,,\\
 r_{\phi^4 D^4}^{(8)} &= -8 c_{\phi\Box}^2 + 2 c_{\phi\Box} c_{\phi D} - \frac{1}{4}c_{\phi D}^2\,,\\
 r_{\phi^4 D^4}^{(10)} &= -c_{\phi D}^2\,,\\
 r_{\phi^4 D^4}^{(11)} &= -16 c_{\phi\Box}^2 + 4 c_{\phi\Box} c_{\phi D}  - \frac{1}{2}c_{\phi D}^2\,.
\end{align}

Adding Eqs.~\eqref{eq:dir} and \eqref{eq:ind}, we obtain the final result in Eq.~\eqref{eq:cephi}.

\bibliographystyle{style} 

\bibliography{refs} 

\end{document}